\documentstyle[aps,prl,epsf,twocolumn,floats]{revtex}
\begin{document}
\draft
\twocolumn[\hsize\textwidth\columnwidth\hsize\csname @twocolumnfalse\endcsname
\title{ Relationship between resistivity and specific heat in a canonical non-magnetic heavy fermion alloy system:  UPt$_{{\bf 5-x}}$Au$_{\bf x}$} 
\author{B. Andraka$^a$, R. Pietri$^a$, S. G. Thomas$^a$, 
G. R. Stewart$^{a,b}$,\\
E. - W. Scheidt$^b$, and T. Schreiner$^b$} 
\address{$^a$Department of Physics, University of Florida,\\
P.O. Box 118440, Gainesville, Florida  32611\\
$^b$University of Augsburg, Memminger Str. 6, 86135 Augsburg, Germany}

\date{\today} 

\maketitle 


\begin{abstract}

UPt$_{5-x}$Au$_x$ alloys form in a single crystal structure, cubic
AuBe$_5$-type, over a wide range of concentrations from $x$ = 0 to at least
$x$ = 2.5. All investigated  alloys, with an exception for $x$ = 2.5, were
non-magnetic. Their electronic  specific heat coefficient $\gamma$ varies from
about 60 ($x$ = 2) to about 700 mJ/mol K$^2$ ($x$ = 1). The electrical
resistivity for all alloys has a Fermi-liquid-like  temperature variation,
$\rho = \rho_o + AT^2$, in the limit of T$\rightarrow$0 K. The coefficient $A$
is  strongly enhanced in the heavy fermion regime in comparison with normal
and  transition metals. It changes from about 0.01 ($x$ = 0) to over 2 $\mu
\Omega$ cm/K$^2$ ($x$ = 1).  $A/\gamma^2$, which has been postulated to have a
universal value for heavy fermions,  varies from about 10$^{-6}$ ($x$ = 0, 0.5)
to 10$^{-5}$  $\mu \Omega$ cm (mol K/mJ)$^2$ ($x > 1.1$),  thus from a value
typical of transition metals to that found for some other  heavy fermion
metals. This ratio is unaffected, or only weakly affected, by  chemical or
crystallographic disorder. It correlates with the paramagnetic 
Curie-Weiss temperature of the high temperature magnetic susceptibility. 

\end{abstract}


\bigskip
]
\narrowtext


Despite more than two decades of research on heavy fermions, the  understanding
of the phenomena remains unsatisfactory, especially for  actinide-based
systems. A notable advance in delineating the origin of the  enhanced
electronic mass in Ce-based systems has been achieved via alloying  studies. 
\cite{1} Such studies are able to show correlations between various low 
temperature characteristics of materials within a single crystal structure.
Few  successful alloying studies involving U-based heavy fermions have been
reported  so far. One of the serious obstacles in such investigations is the
proximity to  magnetism and the robust character of magnetism in U-compounds
and alloys.  Thus, an ideal candidate for such an investigation should  clearly
possess a ground state that is neither magnetic nor superconducting.  To avoid
any  complications due to anisotropy of the electronic properties associated
with the crystal structure, materials crystallizing in cubic structures are
preferred.

There are only few U-alloys and compounds that can be undoubtedly  classified
as heavy fermions and which satisfy these prerequisites. One of them, and
probably the clearest case is UPt$_4$Au. \cite{2} This alloy forms in a cubic 
AuBe$_5$-type crystal structure. Its low temperature specific heat divided by 
temperature is almost 700 mJ/mol K$^2$ at 1 K. The ratio of the magnetic 
susceptibility to the specific heat coefficient at low temperatures and the 
Wilson ratio are small in comparison with all other known heavy fermion
systems.  Equally small ratios have been found only for two heavy fermion
superconductors,  UPt$_3$ and UBe$_{13}$ \cite{3} which lead to speculations
that UPt$_4$Au can be a  heavy fermion superconductor at sufficiently low
temperatures. However,  magnetization measurements \cite{4} performed down to
20 mK have not yielded any  evidence of superconductivity or magnetism. 

It has been reported \cite{5} that UPt$_{5-x}$Au$_x$ alloys form a
single-phase, cubic AuBe$_5$-crystal structure over an extended range of
concentrations $x$ ($x$ = 0 to at least $x$ = 2.5). At the same time, their
electronic properties, like the  low temperature specific heat and magnetic
susceptibility, vary a great deal,  making this pseudobinary alloy system a
unique and convenient system to study  the development of a heavy fermion state
upon varying $x$, and to search for  possible deviations from Landau's
Fermi-liquid theory in the heavy-fermion  regime.

UPt$_{5-x}$Au$_x$ alloys have been prepared by arc melting. All alloys in the 
range $x$ = 0 to 2.5 were single phase according to the X-ray diffraction
analysis.  Based on a previous study, \cite{5} which did not find any
significant dependence of the low temperature properties of these alloys on
annealing, we have not annealed our samples. In our present investigation, we
have used some of those previously investigated unannealed samples.

\begin{figure}
\centerline{\epsfxsize=3.0in \epsfbox {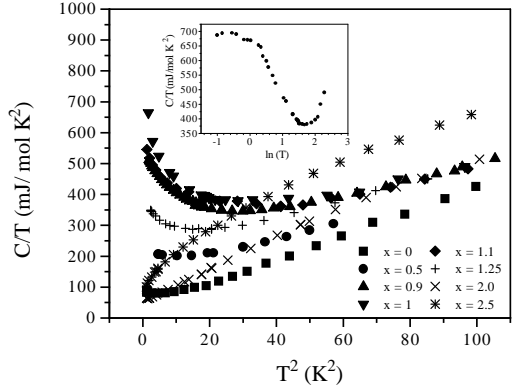}}
\vskip -3cm
\protect \caption{$C/T$ versus $T^2$ for UPt$_{5-x}$Au$_x$ alloys. The
inset shows $C/T$ versus $ln T$ for $x$=1 and temperatures between 0.3 and 10
K.}
\protect \label{f1}
\end{figure}

Specific heat has been measured between about 1 K and 10 K with the exception
for UPt$_4$Au, which has been studied to 0.35 K. The total specific heat
divided by temperature ($C/T$) versus temperature square is shown in
Fig.~\ref{f1} for  eight different alloys corresponding to $x$ between 0 and
2.5.  $C/T$ at the lowest  measured temperature, which we call further
$\gamma$, spans a large range of values. It increases from about 80 mJ/mol
K$^2$ for $x$ = 0 to almost 700 mJ/mol K$^2$ for $x$ = 1, and then drops again
to about 60 mJ/mol K$^2$ for $x$ = 2, thus changing by more than an order of
magnitude. 

A better estimate of $\gamma$ would be $C/T$ at 0 K obtained from the fit of
the low  temperature specific heat data to some theoretical function. This
could be especially important for UPt$_4$Au which exhibits a number of
similarities to extensively studied UCu$_4$Pd \cite{6,7}. In this later
compound, also forming in the AuBe$_5$-type crystal structure, $C/T$ diverges
as T approaches zero. $C/T$ of UPt$_4A$u, between about 1 and 10 K, has also a
quasi-logarithmic temperature dependence if the phonon contribution is
subtracted. However, below 1 K, the $C/T$ data show a clear tendency towards
saturation (and possibly a very weak maximum near 0.6 K which can not be
resolved due to the data scattering). This saturation is shown in the inset to
Fig.~\ref{f1} by expressing the data in the $C/T$ versus ln T format. $C/T$
between 1.1 and 0.36 K varies by less than 5 $\%$ which is less than the
absolute accuracy of the measurement, about 10 $\%$. This result provides some
justification of associating $\gamma$ with the value of $C/T$ at about 1 K.

In agreement with Ref. 8, we have found that the low temperature specific heat
data of UPt$_5$ can be well described by a spin-fluctuation formula. This
formula can still be applied to $x$ = 0.5, but fails to describe the
temperature dependence of $C/T$ for all other investigated concentrations. The
specific heat for these other concentrations can not be associated with either
Kondo-type behavior or, more generally, with any type of behavior corresponding
to a single-impurity scaling, $C(T,T_K) = C(T/T_K)$. \cite{9} Because of almost
identical molecular weights across the series, a small phonon contribution at 1
K is expected to be constant. Attempting to account for the phonon
contribution, we have investigated two non-$f$ AuBe$_5$  homologues: ThPt$_4$Au
and YPt$_4$Au.  \cite{10} The specific heat for both ThPt$_4$Au and  YPt$_4$Au
has a spin-fluctuation temperature dependence with  $\gamma$ values  of 11 and
17 mJ/mol K$^2$, respectively.  Since neither of them contains 5$f$-elements,
the spin-fluctuations have to be associated with 5$d$-electrons. This is an
important observation suggesting that the 5$d$ electrons of Pt can also  be
responsible for the spin-fluctuation term in the specific heat of UPt$_5$, and
that 5$f$-5$d$ hybridization effects can play a significant role in other
UPt$_{5-x}$Au$_x$ alloys, especially in those corresponding to small values of
$x$. A similar conclusion has been reached by authors of Ref. 8, based on a
magnetic susceptibility study.  They have argued that an unusually large
absolute value of the Curie-Weiss temperature ($\sim$ 500 K) in the magnetic
susceptibility of the pure UPt$_5$ compound reflects strong 5$f$-delocalization
effects due to the 5$d$-electrons of Pt.

\begin{figure}[tb]
\protect \centerline{\epsfxsize=3.0in \epsfbox {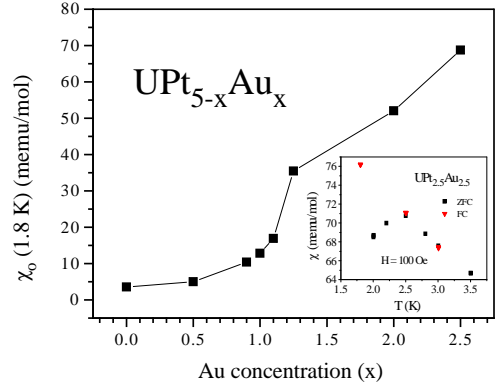}}
\vskip -3cm
\protect \caption{$\chi_o$ versus $x$ for UPt$_{5-x}$Au$_x$. The inset shows 
 zero-field-cooled (ZFC) and field-cooled (FC)   susceptibility ($H = 100$ Oe)
 for UPt$_{2.5}$Au$_{2.5}$.}
\protect \label{f2}
\end{figure}

Magnetic susceptibility at the lowest temperature measured, $\chi_o$ (1.8 K),
shown as a function of $x$ in Fig.~\ref{f2}, does not correlate with $\gamma$
at all.  $\chi_o$ increases monotonically from 3.6 to almost 70 memu/mol
between $x$ = 0 and $x$ = 2.5 (Fig.~\ref{f2}). Thus the ratio $\chi_o/\gamma$
is not constant across the system, as expected from  the single-impurity model,
but increases with $x$. The especially strong increase in $\chi_o$ for $x >
1.25$ prompted us to perform low field magnetization measurements on these
alloys. These measurements for UPt$_{2.5}$Au$_{2.5}$ revealed a maximum at
about 2.5 K  ($T_M$) in the zero field cooled susceptibility (ZFC) and a
discrepancy between  zero-field-cooled and field-cooled (FC) susceptibilities
at low temperatures  (Inset to Fig.~\ref{f2}). No corresponding maximum has
been found in the specific heat data. Similar spin-glass-like anomalies have
been observed in other $\gamma$-enhanced U- or Ce-based systems, with chemical
formulas  of the form M(A$_x$B$_{1-x}$), near $x$ = 0.5; where M is either U or
Ce, and A and B  are normal or transition metals.  \cite{11} Although there is
no consensus as to the  nature of the ground state (or different ground states)
in these alloys, it is  believed that non-magnetic atom disorder (NMAD) is
responsible for their spin-glass-like properties.  Low field magnetization
measurements performed on  all other alloys, including UPt$_2$Au$_3$, have not
detected any anomalies, although  small discrepancies (up to 5$\%$) between ZFC
and FC susceptibility have been found for $x$ = 2 below 3 K.  The maximum
value  of $T_M$ occurring for $x \simeq$ 2.5 in the UPt$_{5-x}$Au$_x$ system is
consistent with  NMAD reaching its maximum near $x$ = 2.5.  Magnetic
measurements indicate further that UPt$_{5-x}$Au$_x$ alloys are not completely
free from complications associated with the nearness to magnetism. 
Nevertheless, such complications can be safely ruled out in the range of $x$
where $\gamma$ undergoes especially spectacular changes ($0 \leq x \leq 1.25$).

\begin{figure}[tb]
\protect \centerline{\epsfxsize=3.0in \epsfbox {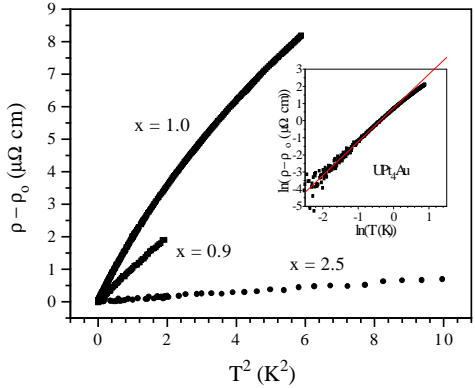}}
\vskip -3cm
\protect \caption{$\rho - \rho_o$ versus $T^2$ for $x$ = 0.9, 1, and 1.25. The inset 
 shows the zero-temperature  limit of the resistivity for UPt$_4$Au on a
 logarithmic scale.}
\protect \label{f3}
\end{figure}

Before addressing our resistivity results we point out the important issue of
preferential site occupancy. Strong variation of the thermodynamic low
temperature properties of UPt$_{5-x}$Au$_x$ near $x$=1 and their extremal
values for $x$=1 have been attributed to preferential atomic occupancy of non-U
sites \cite{2,5}. Pt would preferentially occupy trigonal symmetry 16e sites
while Au would go on cubic 4c sites. Additional, indirect support for this
explanation is given by thermodynamic investigations of the aforementioned
U-alloy system UCu$_{5-x}$Pd$_x$, existing in the same crystal structure. The
concentration corresponding to $x$=1 also exhibits extremal values of its low
temperature properties in comparison to other alloys belonging to
UCu$_{5-x}$Pd$_x$ system. Note that the smaller atoms (Cu, Pt) would
preferentially occupy majority 16e sites, while larger atoms would occupy 4c
sites in both alloy systems. Previous attempts using X-ray diffraction and
resistivity studies to clarify this issue for UPt$_{5-x}$Au$_x$ were
inconclusive \cite{5}. However, a recent elastic neutron scattering study
\cite{12} on the related UCu$_{5-x}$Pd$_x$ has confirmed the preferential non-U
site occupancy. Thus, a similar preferential site occupancy and chemical
ordering of $x$=1 is highly probable for UPt$_{5-x}$Au$_x$.

This resistivity for all investigated alloys has a Fermi-liquid-like
temperature dependence, $\rho = \rho_o + AT^2$, at sufficiently low
temperatures. This   variation is shown in Fig.~\ref{f3} for $x$ = 0.9, 1, and
2.5 in the form of $\rho -  \rho_o$ versus $T^2$. The residual resistivity
$\rho_o$, obtained on samples derived from a single batch of U, increases with
$x$ from about 25  $\mu \Omega$ cm for $x$ = 0 \cite{8} to about 165 $\mu
\Omega$ cm for $x$ = 2.5. Thus in agreement with the previous study, our
resistivity results do not directly support the notion of a chemical ordering
near $x$=1; i.e., no discernable dip in $\rho_o$ is observed for this
concentration. We have noticed some variation of $\rho_o$ between samples
obtained from U-batches having different chemical purity. One of our UPt$_4$Au
samples had $\rho_o \simeq 40 \mu \Omega$ cm, the other one had $\rho_o \simeq
80 \mu \Omega$ cm. Interestingly, the temperature square coefficient of the
resistivity $A$, discussed next, was identical for both $x$=1 samples, within
the experimental uncertainty (10 $\%$).

\begin{figure}[tb]
\protect \centerline{\epsfxsize=3.0in \epsfbox {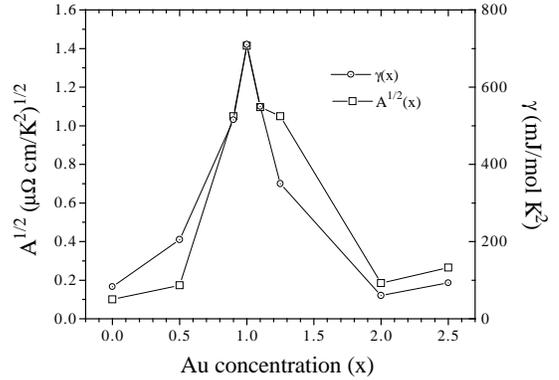}}
\vskip -3cm
\protect \caption{$A^{1/2}$ and $\gamma$ versus $x$ for UPt$_{5-x}$Au$_x$.}
\protect \label{f4}
\end{figure}

\begin{figure}[tb]
\protect \centerline{\epsfxsize=3.0in \epsfbox {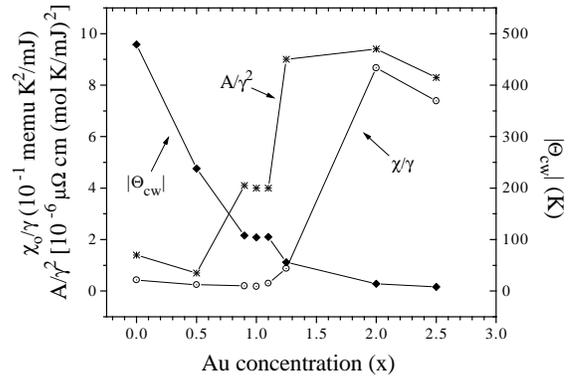}}
\vskip -3cm
\protect \caption{$A/\gamma^2$, $\chi_o/\gamma$, and $|\Theta_{\mbox{{\tiny CW}}}|$ 
 versus $x$.}
\protect \label{f5}
\end{figure}

The most striking feature of Fig.~\ref{f3} is a large variation, from
concentration to concentration, in the slope of the $\rho - \rho_o$ versus
$T^2$.  This slope corresponds to the $A$ coefficient. In the investigated
range of concentrations, $A$ changes, in a non-monotonic fashion, from its
lowest value of about 0.01 $\mu \Omega$  cm/K$^2$ for $x$ = 0 to over 2 $\mu
\Omega$ cm/K$^2$ for $x$ = 1, thus by a  factor larger than 200. The $A$
coefficient fairly well tracks the trends for the changes in $\gamma$,  as
demonstrated in Fig.~\ref{f4}. The format of this figure, $A^{1/2}$ and 
$\gamma$ versus $x$, has  been chosen based on the observation of Kadowaki and
Woods, \cite{13} that for a large  number of heavy fermion compounds $A \simeq
10^{-5} \gamma^2$, where $A$ is expressed in $\mu \Omega$ cm/K$^2$ and $\gamma$
in mJ/mol K$^2$.  A similar relationship has been found in transition metals,
but with the average ratio $A/\gamma^2$ about 20 times smaller than that in
heavy fermion compounds.  This proportionality between $A$ and $\gamma^2$ in
transition metals has been accounted  for theoretically \cite{14} in terms of
Baber's model \cite{15} describing electron-electron scattering.  Also, the
electron-electron scattering is believed to be the source  of the enhanced
values of $A$ in heavy fermion systems \cite{16,17}.  However, the large ratio
$A/\gamma^2$, in comparison with transition metals, has not yet been fully
explained.  Furthermore, the universality of this ratio has not been
established.  The ratio of $A$ and $\gamma^2$ is not constant across the
UPt$_{5-x}$Au$_x$ series (Fig.~\ref{f5}).  This ratio is of order 10$^{-6}$
$\mu \Omega$ cm (mol K/mJ)$^2$ only for the pure UPt$_5$  compound and the $x$
= 0.5 alloy, thus within the range of values corresponding to many transition
metals.  $A/\gamma^2$ grows to  $4 \times 10^{-6}$  $\mu \Omega$ cm (mol
K/mJ)$^2$ for $x$ = 0.9, 1, and 1.1, and reaches a value close to the claimed
universal  ratio for heavy fermion compounds for $x$ = 1.25, 2, and 2.5.  Our
results confirm the correlation between $A$ and $\gamma^2$, but they  also
indicate that the ratio of these two quantities can vary by more than a decade,
in disagreement with recent theoretical predictions \cite{17}.  To the best of
our knowledge, this is the first observation of such a correlation in an alloy
system over a large range of concentrations, as opposed to a previous study on 
pure compounds possessing full translational symmetry. \cite{13}

It has been postulated \cite{16} that disorder can influence $A/\gamma^2$
values.  Both, the ratio $A/\gamma^2$ and the amount of crystallographic
disorder, in general, increase with $x$ in our alloy system, therefore implying
that the disorder increases this ratio.  However, such a possibility is
unlikely considering that the two UPt$_4$Au samples studied showed
significantly different residual resistivities,  as discussed earlier, and yet
had identical values of $A/\gamma^2$ within the accuracy of our measurement.
Thus crystallographic or atomic disorder seems to play a rather minor, if any,
role in the evolution of $A/\gamma^2$ throughout the series.

It has been also argued, in the framework of the spin-fluctuation  model,
\cite{17} that this ratio, which should be almost universal for heavy fermions,
can be enhanced in the proximity to a magnetic instability.  According to
experimental studies \cite{3}, the  Wilson ratio ($\chi_o/\gamma
\mu_{\mbox{{\tiny eff}}}^2$) can be used as an indication of such a proximity
to magnetism.  Since $\mu_{\mbox{{\tiny eff}}}$ extracted from the high
temperature susceptibility is almost constant across this pseudobinary alloy
system, one may use the $\chi_o/\gamma$ ratio instead.  $\chi_o/\gamma$ versus
$x$ is shown in Fig.~\ref{f5} together with corresponding $A/\gamma^2$ values. 
Although both ratios increase above $x$ = 1.25, there is no correlation between
these two quantities for $x < 2$.  A somewhat better correlation is observed
between $A/\gamma^2$  and the  absolute value of the  Curie-Weiss temperature
($|\Theta_{\mbox{{\tiny CW}}}|$). A decrease of  $|\Theta_{\mbox{{\tiny
CW}}}|$  corresponds to an increase of $A/\gamma^2$. In particular,
$|\Theta_{\mbox{{\tiny CW}}}|$ and  $A/\gamma^2$ are constant for $x$ = 0.9, 1,
and 1.1, thus for a concentration range corresponding to large changes  of
$\gamma$.  It is reasonable to associate large values of $|\Theta_{\mbox{{\tiny
CW}}}|$ in this alloy system, in agreement with Ref. 6 and also with our
specific heat data, with strong  5$f$-5$d$ hybridization, and consequently with
a large degree of delocalization of 5$f$  electrons.  Alloys for which we
expect especially strong 5$f$-5$d$ hybridization, as  evidenced by large
paramagnetic Curie-Weiss temperatures ($x\sim 0$), have  $A/\gamma^2$  values
in the range corresponding to transition metals.  Upon a decrease of
$|\Theta_{\mbox{{\tiny CW}}}|$, $A/\gamma^2$  increases and reaches a value
close to 10$^{-5}$ $\mu \Omega$ cm (mol K/mJ)$^2$, found in many other heavy
fermion  systems.

The discussed Fermi-liquid-like temperature dependence in the resistivity of
UPt$_{5-x}$Au$_x$ alloys is observed only at the lowest temperatures, and   the
range of temperatures in which it holds varies with $x$. The upper temperature
limit of this variation seems to correlate with the inverse of $\gamma$; e.g.,
it is  about 1 K for $x$ = 0.9, 0.7 K for $x$ = 1.0, 4K for $x$ = 2.0, and
about 3 K for  $x$ = 2.5.  This observation is rather suprising since it is
generally believed that the  $\rho = \rho_o + AT^2$ temperature dependence  is
a property of the so-called coherence regime, which is a  consequence of a
translational periodicity of the system.  Our results do not preclude the
possibility that replacing Pt by Au, leading to the distraction of  this
symmetry, affects somehow the upper temperature limit of the discussed
temperature variation, but it does not seem to be a dominating cause.

Finally, our results have implications on the possible explanation of the
origin of the non-fermi-liquid-like low temperature properties for the
extensively studied UCu$_{5-x}$Pd$_x$ alloys \cite{6,7}. One of the leading
explanations \cite{18} is based on the Kondo-disorder idea \cite{19}. Despite
the obvious ligand disorder also in UPt$_{5-x}$Au$_x$ alloys, their properties
stay fermi-liquid-like over a wide range of $x$. This observation imposes
severe limitations on the applicability of the Kondo-disorder theory to
UCu$_4$Pd. 

Work at Florida was supported by National Science Foundation, grant  number
DMR-9400755 and Department of Energy, grant number DE-FG05-86ER45268.




\begin{references}
\bibitem{9}This scaling relation, $C(T,T_K) = C(T/T_K)$, provides a convenient
method to  determine the characteristic temperature $T_K$, which is inversely
proportional to  $\gamma$.  However, the specific heat data for
UPt$_{5-x}$Au$_x$ alloys can not  be superimposed  on a single curve by
appropriate scaling of the temperature axis. Both $C/T$  and $C$ for $x$ = 0.9,
1.0, and 1.1 overlap above 8 K, but span a large range  of values at 1 K.
\end{references}
\end{document}